\documentclass[a4paper]{article}
\usepackage{amsmath,pstricks}
\newcommand{\scr}{\scriptscriptstyle}
\newcommand{\eqa}{\begin{eqnarray}}
\newcommand{\neqa}{\end{eqnarray}}
\newcommand{\equ}{\begin{equation}}
\newcommand{\nequ}{\end{equation}}

\newcommand{\no}{\nonumber\\}
\newcommand{\Ref}[1]{(\ref{#1})}

\begin{document}

\title{\bf{\LARGE The complete LQG propagator: \\ 
II. Asymptotic behavior of the vertex}} 
\author{\large
Emanuele Alesci${}^{ab}$, Carlo Rovelli${}^b$
 \\[3mm]
\em
\normalsize
{${}^a$Dipartimento di Fisica Universit\`a di Roma Tre, I-00146 Roma EU}
\\ \em\normalsize{${}^b$Centre de Physique Th\'eorique de Luminy%
\footnote{Unit\'e mixte de recherche (UMR 6207) du CNRS et des
Universit\'es de Provence (Aix-Marseille I), de la Mediterran\'ee
(Aix-Marseille II) et du Sud (Toulon-Var); laboratoire affili\'e \`a
la FRUMAM (FR 2291).}, Universit\'e de la M\'editerran\'ee, F-13288
Marseille EU}}

\date{\small\today}
\maketitle \vspace{-.6cm}

\begin{abstract}

\noindent In a previous article we have show that there are 
difficulties in obtaining the correct graviton 
propagator from the loop-quantum-gravity dynamics defined by the 
Barrett-Crane vertex amplitude. Here we show that a vertex amplitude 
that depends 
nontrivially on the intertwiners can yield the correct propagator.  We give 
an explicit example of asymptotic behavior of a vertex amplitude 
that gives the correct full graviton propagator in the large distance 
limit.
\end{abstract}
\section{Introduction}

A technique for computing $n$-point functions in a background-independent
context has been recently introduced \cite{scattering1, scattering3}
and developed \cite{Livine:2006it}. 
Using this technique, we have found in a previous paper \cite{I} that the 
definition of the dynamics of loop quantum gravity (LQG) by means of the
Barrett-Crane (BC) spinfoam vertex \cite{Barrett:1997gw} \emph{fails} to give 
the correct  tensorial structure of the graviton 
propagator in the large-distance limit.  The natural question is
whether this is an intrinsic difficulty of the background-independent
loop and spinfoam formalism, or whether it is a specific difficulty of the
BC vertex.  
Here we show that the answer is the second.  We do so by  explicitly 
exhibiting a vertex amplitude $W$ that yields the correct propagator in 
the large distance limit.   We have no claim that this vertex amplitude is
physically correct. In fact, it is a rather artificial
object, chosen by simply taking the asymptotic form of the BC
vertex, and correcting the detail for which the BC vertex fails to work.
Thus,  $W$ has at best an interest in the asymptotic region.  But its 
existence shows that the  background-independent loop and spinfoam 
formalism, \emph{can} yield the full tensorial structure of the perturbative
$n$-point functions. 

Furthermore, the properties of $W$ give some indications 
on the asymptotic that the dynamics can have, if it has to
yield the correct low energy limit.  The detail of the BC vertex that 
needs to be corrected turns out to 
be a \emph{phase} in the intertwiner variables.  A posteriori, the
need for this phase appears pretty obvious
on physical grounds, as we shall discuss in detail.  This might provide
a useful indication for selecting a definition of the dynamics alternative
to the one provided by the BC vertex.   While the
BC vertex is defined by the $SO(4)$ Wigner 10j  symbol, 
an alternative
vertex given by the square of an $SU(2)$ Wigner 15j 
symbol has been introduced recently \cite{EPR}. This vertex 
can be derived also using coherent states techniques, and can be extended to the 
Lorentzian case and to arbitrary values of the Immirzi parameter
\cite{LS}.    It would be very interesting to see whether the asymptotics of this vertex exhibit
the phase dependence that we find here to be required for the low energy limit. 

In Section II we introduce the vertex $W$ and we give a simple explanation
of the reason why the additional phase is
needed.  In the rest of the paper we prove that $W$  yields  the correct 
full tensorial structure of the propagator.   In developing this calculation we 
have stumbled upon an unexpected result that indicates that the state 
used in  \cite{I} is too symmetric.  This does not affect the results
of  \cite{I}, but forces us to reconsider the definition of the state.  In section III, we 
discuss this issue in detail and
give the appropriate boundary state.  In Section IV we compute the propagator, and
in Section V we compare it with the one computed in linearized quantum general 
relativity.  

This paper is not self-contained.  It is based on the paper  \cite{I}, where all
relevant definitions are given.   For an introduction to the formalism we use,
see \cite{scattering3}; for a general introduction to background independent
loop quantum gravity \cite{lqg2}, see \cite{lqg}.

\section{The vertex and its phase}

Following \cite{scattering1,scattering3}, the 
graviton propagator can be computed in a background 
independent context as the scalar product 
\begin{equation}
{\mathbf G} _{{\mathbf q}\, n,m}^{\scriptscriptstyle ij,kl} = \langle W | \big(E^{\scriptscriptstyle(i)}_n \cdot E^{\scriptscriptstyle(j)}_n-n_n^{\scriptscriptstyle(i)}\cdot n_n^{\scriptscriptstyle(j)}\big)
\big(E^{\scriptscriptstyle(k)}_m  \cdot E^{\scriptscriptstyle(l)}_m-n_m^{\scriptscriptstyle(k)}\cdot n^{\scriptscriptstyle(l)}_m\big) |\Psi_{\mathbf q} \rangle.
\label{partenza1}
\end{equation}
Here $ \langle W |$ is the boundary functional, which can be intuitively understood as the 
path integral of the Einstein-Hilbert action on a finite spacetime region $\cal R$, with given
boundary configuration.    The indices $i,j,k,l,m,n,...$ run over the values $1,...,5$ and label the tetrahedra of a 4-simplex.  The operator $E_n^{\scriptscriptstyle(i)}$ (denoted $E_n^{\scriptscriptstyle(ni)}$ in \cite{I}) is the triad operator at the points $n$, contracted with (test) one-forms $n_n^{\scriptscriptstyle(i)}$  (denoted $n^{ni}$ in \cite{I}) at the same point.  $|\Psi_{\mathbf q} \rangle$
is a state on the boundary of $\cal R$, picked on a given classical boundary (extinsic and 
extrinsic)  geometry $\mathbf q$.

Fixing such a boundary geometry is equivalent to fixing a background metric $g$ in the 
interior, where $g$ is the solution of the Einstein equations with boundary data $\mathbf q$. 
The existence of such a background metric is part of the 
definition of the propagator,
which is a measure of fluctuations around a given background. Criticisms to the
approach of  \cite{scattering1,scattering3} have been raised on the
ground that a propagator makes no sense in a background independent context, 
because it is a quantity that depends on a background geometry. These criticisms
follow from a misunderstanding of this point. The information about the background 
over which the propagator is defined is in the boundary state, via ${\mathbf q}$.

We are interested in the value of \Ref{partenza1} to first order in the GFT expansion 
parameter $\lambda$, and in the limit in which the boundary surface (whose size is 
determined by  $\mathbf q$) is large. On the physical interpretation of this limit, see 
\cite{EPR}.  To first order, the leading contribution to 
$W$ has support only on spin networks with a 4-simplex graph. If 
${\mathbf j}=(j_{nm})$ and ${\mathbf i}=(i_n)$
are, respectively, the ten spins and the five intertwiners that color this graph, then in this
approximation (\ref{partenza1}) reads
\begin{equation}
{\mathbf G} _{{\mathbf q}\, n,m}^{\scriptscriptstyle ij,kl} 
 = \sum_{{\mathbf j}, {\mathbf i}} W({\mathbf j}, {\mathbf i}) \big(E^{\scriptscriptstyle(i)}_n \cdot E^{\scriptscriptstyle(j)}_n-n_n^{\scriptscriptstyle(i)}\cdot n_n^{\scriptscriptstyle(j)}\big) 
 \big(E^{\scriptscriptstyle(k)}_m  \cdot E^{\scriptscriptstyle(l)}_m-n_m^{\scriptscriptstyle(k)}\cdot n^{\scriptscriptstyle(l)}_m\big)
 \Psi({\mathbf j}, {\mathbf i}).
\label{partenza} 
\end{equation}
To this order, $W$ is just determined by the amplitude of a single vertex. In \cite{scattering1,scattering3, I}, (a suitable adjustment of) the BC vertex was chosen
for $W$. The propagator depends only on the asymptotic behavior of the vertex.
This has the structure \cite{Barrett:1998gs}
\begin{equation}
	W_{BC}(\mathbf{j})\sim
	e^{\frac{i}{2}(\delta{\mathbf{j}} G \delta {\mathbf{j}})} e^{i\Phi \cdot \delta{\mathbf{j}}} +e^{-\frac{i}{2}(\delta {\mathbf{j}}G \delta {\mathbf{j}})} e^{-i\Phi\cdot \delta {\mathbf{j}}}
	\label{BC},
\end{equation}
where $G$ is the $10\times10$ matrix given by the second derivatives of the 4d Regge action around
the symmetric state, $\delta  {\mathbf{j}}$ is the difference between the ten spins ${\mathbf{j}}$  
and their  background value $j_0$, and $\Phi$ is a 10d vector with all equal components, which were shown in \cite{scattering1,scattering3} to precisely match those determined by the background extrinsic curvature.  The diagonal components of the propagator determined
by  (\ref{partenza1}) turn out to be correct at first  \cite{scattering1} and second  
 \cite{scattering3} order, but the nondiagonal components fail to do so \cite{I}. 
 
Here we make a different choice for $W$.   We choose a vertex $W$ with an 
asymptotic form that includes a gaussian intertwiner-intertwiner and spin-intertwiner  
dependence, and --most crucially-- a phase dependence on the intertwiner variables. 
To write this, introduce a 15d vector  $\delta \mathbf{I}=(\delta {\mathbf{j}}, \delta {\mathbf{i}})$,
where $\delta  {\mathbf{i}}$ is the difference between the five intertwiners ${\mathbf{i}}$  
and their  background value $i_0$.   Explicitly, 
 $\delta I_\alpha=(\delta j_{nm}, \delta i_n)=(j_{nm}-j_0,  i_n-i_0)$, where $\alpha=(nm, n)$.
And consider the state 
\begin{equation}
	W(\mathbf{j},\mathbf{i}) = 
	e^{\frac{i}{2}(\delta{\mathbf{I}} G \delta {\mathbf{I}})} e^{i\phi \cdot \delta{\mathbf{I}}} +e^{-\frac{i}{2}(\delta {\mathbf{I}} G \delta {\mathbf{I}})} e^{-i\phi\cdot \delta {\mathbf{I}}}.
	\label{caso1}
\end{equation}
Here $G$ is now a $15\times15$  matrix and $\phi=(\phi_{nm},\phi_n)$ is a 15d vector. Its 10 spin components
$\phi_{nm}$ just reproduce the spin phase dependence of \Ref{BC}; while its five intertwiner components are equal and we fix them to have value 
\begin{equation}
\phi_n=\frac{\pi}{2}.  
\end{equation}
This phase dependence is the crucial detail that makes the calculation work. 

Let us  illustrate upfront the reason why this additional phase cures the problems that appeared with the BC vertex.  The boundary state must have an intertwiner dependence, in order to have the
correct semiclassical value of the mean values of the angles between the faces of the boundary tetrahedra. The mean value of an intertwiner variable $i_n$ --namely of the virtual link
of the intertwiner in a given pairing-- must have a certain value $i_0$. For this, it is sufficient, say,
that the state be a gaussian around $i_0$.  However, in quantum
geometry the different angles of a tertrahedron do not commute 
\cite{tetraedro}. Therefore a state with
a behavior like $exp\{-(i_n-i_0)^2\}$ will be peaked on the virtual spin $i_n$ in one pairing,
but {\em it will not be peaked in the virtual spin in a different pairing}.  Therefore, the other angles
of the tetrahedron will not be peaked on the correct semiclassical value.   We can of course
write a gaussian which is peaked on a variable as well as on another, non-commuting, 
variable.   For instance, a standard Schr\"odinger wave packet $\psi(x)=exp\{-\frac{(x-x_0)^2}{2}\sigma+ip_0x\}$ is peaked on position as well as momentum.   But in order to do so, we 
must have a phase dependence on the $x$.  Similarly, the boundary state needs a
phase dependence on the intertwiner variable $i_n$, in order to be peaked on all angles.   
As shown in \cite{sc}, the correct value for this is $exp\{i\frac{\pi}{2}i_n \}$.
Now, the general mechanism through which the dynamical kernel reproduces the 
semiclassical dynamics in quantum mechanics is
the cancellation of the phases between the propagation kernel and the boundary state. If
this does not happens, the rapidly oscillating phases suppresses the amplitude. 
For instance, in the non-relativistic quantum mechanics of a free particle, 
the propagation kernel $K(x,y)$ in a time  $t$ has a phase dependence from small fluctuations 
$\delta x=x-x_0$ and $\delta y=y-y_0$ of the form 
\begin{equation}
K(x_0+\delta x,y_0+\delta y)=\langle x_0+\delta x |e^{-\frac{i}{\hbar}\frac{p^2}{2m}t}| y_0+\delta y \rangle \sim C \ e^{-ip_0\delta x}\  e^{ip_0\delta y}. 
\end{equation}
where $p_0=m(y_0-x_0)/t$. This phase precisely  cancels the phase of an initial and final wave 
packets $\psi_i$ and $\psi_f $ centered on $x_0$ and $y_0$, if these have the correct momentum.  
That is 
\begin{equation}
\langle \psi_f |e^{-\frac{i}\hbar Ht}| \psi_i \rangle
= \int dx \int dy\  e^{-\frac{(x-x_0)^2}{2\sigma}-\frac{i}{\hbar}p_f x} \ 
K(x,y)\  e^{-\frac{(y-y_0)^2}{2\sigma}+\frac{i}{\hbar}p_i y}
\end{equation}
is suppressed by the oscillating phases unless $p_i=p_f=p_0$.  This is the standard mechanism
through which quantum theory reproduces the (semi-)classical behavior.
In quantum gravity, it is reasonable to expect the same to happen if we have to recover the Einstein equations in the semiclassical limit.  That is, the propagation kernel $W$, must have a phase dependence that matches the one in a semiclassical boundary state. This is precisely the role of the phase $exp\{i\frac{\pi}{2}i_n \}$ that we have included in \Ref{caso1}. 

In the rest of the paper we show that a vertex amplitude that has the phase dependence
as above can reproduce the tensorial structure of the graviton propagator.   First, however,
we must improve the definition of the vertex given above, and correct a problem with the definition of the state in \cite{I}.

\section{Boundary state and symmetry}

Following \cite{scattering1,scattering3}, we consider a boundary state defined as a 
gaussian wave packet, centered on the values determined
by the background geometry $\mathbf q$. Here  
 \begin{equation}
\Phi_{\mathbf q}({\mathbf j},{\mathbf i}) = C\ 
 e^{-\frac{1}{2j_0}(\delta {\mathbf I}A\delta {\mathbf I})
+  i \phi \cdot \delta {\mathbf I}}.
\label{coefficientistato}
\end{equation} 
Where $A$ is a $15\times15$ matrix and the normalization factor $C$ is determined by $\langle W |\Phi_{\mathbf q}\rangle=1$.  The spin phase coefficients are fixed by the background extrinsic geometry \cite{scattering1}. The intertwiner phase coefficients are fixed by requirement that the state remain peaked after a change of pairing to the value $\phi_n=\pi/2$. \cite{I,sc} \   

At each node $n$ we have three possible pairings, that we denote as $x_n$, $y_n$ and $z_n$.
For instance, at the node 5, let $x_5=\{(12)(34)\}, y_5=\{(13)(24)\}, z_5=\{(14)(23)\}$, and
denote $i_{x_5}=i_{\scriptscriptstyle\{(12)(34)\}}$ the intertwiner in the pairing $x_5$, and so on. 
The vertex (\ref{caso1}) and the state \Ref{coefficientistato} are written in terms of the intertwiner variable $i_n$, which is the virtual link  of the node $n$ {\em in one chosen pairing}. 
Because of this, the definition of these states depend on the pairing chosen.  
It follows that  the vertex and the state  do not have 
the full symmetry of the 4-simplex. The corresponding propagator turn out not to be
invariant under $SO(4)$, as it should in the euclidean theory.
In \cite{I},  a simple strategy was adopted in order to overcome this difficulty:  sum over the 
three pairings at each of the five nodes.   The state was defined as 
\begin{equation}
|\Psi_{\mathbf q}\rangle=\sum_{m_n}\  \sum_{\hspace{1em} \mathbf{j},\mathbf{i}_{m_n}}\ 
\Phi_{\mathbf q}(\mathbf{j},\mathbf{i}_{m_n})\ 
|\mathbf{j},\mathbf{i}_{m_n} \rangle,
\label{wrong}
\end{equation}
where $m_n=x,y,z$ for each node $n$. 
This sum implements the full symmetry of the 4-simplex. Summing over the three bases removes the basis dependence.

In developing the calculations presented in the present paper, at first we 
adopted this same strategy.  To our surprise, nothing worked, and something quite
funny happened: the dependence on the intertwiner variables $i_n$
misteriously cancelled out in all components of the propagator! 

The solution of the puzzle was to realize that to sum over the three basis with a correlation matrix A does 
implement the symmetry of the 4-simplex, but not just this symmetry. It implements
a larger symmetry, that has the effect of cancelling the intertwiner dependence.
Geometrically, this additional symmetry can be viewed as an \emph{independent}
rotation of each of the five tetrahedra forming the boundary of the 4-simplex. 

To understand what happens, consider for instance the correlation 
$\langle j_{12}  i_{x_5} \rangle $ between the spin $j_{12}$ which is the quantum 
number of the area of a triangle, and the intertwiner $i_{x_5}$, which is the
quantum number of the angle $\theta_{12}$ between the faces 2 and 3 of the tetrahedron 5. 
More precisely,  $i_{x_5}$ is the eigenvalue 
of the quantity $A_2^2+A_3^2+A_2A_3\cos(\theta_{12})$, where $A_i$ is the area of the face $i$
of the tetrahedron 5.
Now, if the state is summed over pairings, then it does not distinguish pairings, hence
\begin{equation}
\langle  j_{12} i_{x_5}  \rangle= \frac13\left(
\langle  j_{12} i_{x_5}  \rangle+
\langle  j_{12} i_{y_5}  \rangle+
\langle  j_{12} i_{z_5}  \rangle
\right).
\end{equation}
That is 
\begin{equation}
\langle  j_{12} i_{x_5}  \rangle= \frac13
\langle  j_{12} \left(3A_1^2+A_2^2+A_3^2+A_4^2+A_1A_2\cos(\theta_{12})+
A_1A_3\cos(\theta_{13})+A_1A_4\cos(\theta_{14})\right)  \rangle.
 \label{gulp}
\end{equation}
But let $n_i$, $i=1,...,4$ be the normal to the face $i$ of the tetrahedron 5, with length
$| n_i|=A_i$.   The closure relation reads 
\begin{equation}
\sum_{i=1,4} n_i=0.
\end{equation}
Taking the scalar product with $n_1$ gives
\begin{equation}
A_1^2+A_1A_2\cos(\theta_{12})+A_1A_3\cos(\theta_{13})+A_1A_4\cos(\theta_{14})=0.
\end{equation}
It follows from this equation and (\ref{gulp}) that 
\begin{equation}
\langle  j_{12} i_{x_5}  \rangle
= \frac13  \langle j_{12}  (2A_1^2+A_2^2+A_3^2+A_4^2) \rangle
= \frac13  (2\langle j_{12}  j_{15} \rangle
+\langle j_{12}  j_{25} \rangle
+\langle j_{12}  j_{35} \rangle
+\langle j_{12}  j_{45} \rangle).
\end{equation}
That is, the spin-intertwiner correlations are just functions of the spin-spin correlations for a state with this symmetry!  The intertwiner dependence drops out! 
This means that the propagator is completely unaffected from the correlations involving 
the intertwiners. It then turns out that the sole spin-spin correlations in the state are not
sufficient to give the full tensorial structure of the propagator.

The solution of the difficulty is just to choose a boundary state and a kernel $W$ that do not have 
the extra symmetry.   The simplest possibility is to choose an abitrary pairing, and then to symmetrize {\em only} under the symmetries of the four-simplex.   These are generated by the $5!$ permutations $\sigma$ of the five vertices of the four-simplex. A permutation 
$\sigma:\{1,2,3,4,5\}\rightarrow\{\sigma(1),\sigma(2),\sigma(3),\sigma(4),\sigma(5)\}$ 
acts naturally on the  boundary states
\begin{equation}
\sigma | j_{nm}, i_{x_n}\rangle = | j_{\sigma(n)\sigma(m)}, i_{\sigma(x_n)}\rangle
\end{equation}
where the action $\sigma(x_n)$ of the permutation on a node is defined by 
\begin{equation}
\sigma(\{(ab)(cd)_n\})=\{(\sigma(a)\sigma(b))(\sigma(c)\sigma(d))_{\sigma(n)}\}
\end{equation}
and can therefore change the original pairing at the node. 

We therefore define the boundary state by replacing \Ref{wrong} with  
\begin{equation}
\begin{split}
|\Psi_{\mathbf q}\rangle
= \sum_{\sigma}\  \sigma |\Phi_{\mathbf q}\rangle
= \sum_{\sigma}\ \sum_{{\mathbf j},{\mathbf i}}\  \Phi_{\mathbf q}({\mathbf j},{\mathbf i})\ \ 
\sigma |{\mathbf j},{\mathbf i}\rangle. 
\end{split}
  \label{psiB}
\end{equation}
This modification of the boundary state does not affect the conclusions of the paper 
\cite{I}.  Similarly, we pose 
\begin{equation}
 |W\rangle = \sum_{\sigma}\  \sum_{{\mathbf j},{\mathbf i}} \ W({\mathbf j},{\mathbf i})
\ \ \sigma | {\mathbf j},{\mathbf i}\rangle. 
  \label{psiW}
\end{equation}

Before beginning the actual calculation of the propagator, consider what happens by
contracting the vertex amplitude with the boundary state.  We have the double sum over permutations
\begin{equation}
 \langle W|\Psi \rangle = \sum_{\sigma\sigma'}\Big(\sum_{{\mathbf j}{\mathbf i}{\mathbf j}'{\mathbf i}'} \ \overline{W({\mathbf j},{\mathbf i})}
\Phi({\mathbf j}',{\mathbf i}')
\  \langle   \sigma({\mathbf j},{\mathbf i})  | \sigma'({\mathbf j}',{\mathbf i}')\rangle \Big). 
 \label{WPhi}
\end{equation}
The scalar product is 
\begin{equation}
\langle  {\mathbf j},{\mathbf i}  | {\mathbf j}',{\mathbf i}'\rangle= \delta_{{\mathbf j},{\mathbf j}'}
\ \prod_n \   \langle i_n | i'_n \rangle, 
 \label{sp}
\end{equation}
where $ \langle i_n | i'_n \rangle$ is $\delta_{i_n,i_n'}$ if the two intertwiners are written in the same basis, and is  the matrix of the change of basis, namely a $6j$-symbol, otherwise. Now, it was observed in \cite{I} that if one of these $6j$-symbols enters in a sum like  \Ref{WPhi} then 
the sum is suppressed in the large $j_0$ limit, because the $6j$-symbol contains a rapidly oscillating factor which is not compensated.   Hence, in this limit we can effectively rewrite  \Ref{WPhi} in the
form
\begin{equation}
 \langle W|\Psi \rangle = \sum_{\sigma\sigma'}\Big(\sum_{{\mathbf j}{\mathbf i}{\mathbf j}'{\mathbf i}'} \ \overline{W({\mathbf j},{\mathbf i})}
\Phi({\mathbf j}',{\mathbf i}')\ \delta_{\sigma{\mathbf j},\sigma'{\mathbf j}'} \ \delta_{\sigma{\mathbf i},\sigma'{\mathbf i}'} \Big), 
 \label{WPhi2}
\end{equation}
where the second delta vanishes unless the two intertwiners have the same value and are written in the same basis. Up to accidental symmetry factors that we absorb in the state, we can then rewrite the scalar product in the form
\begin{equation}
 \langle W|\Psi \rangle = \sum_{\sigma}\Big(\sum_{{\mathbf j}{\mathbf i}} \ \overline{W({\mathbf j},{\mathbf i})}
\Phi({\mathbf j},{\mathbf i})\ \Big) =  5! \sum_{{\mathbf j}{\mathbf i}} \ \overline{W({\mathbf j},{\mathbf i})}
\Phi({\mathbf j},{\mathbf i}). 
 \label{WPhi3}
\end{equation}
We shall see that a similar simplification happens in the calculation of the matrix elements of the propagator. 

\section{The propagator}

Let us begin by recalling the action of the grasping operators.  This was computed in \cite{I}, to which we refer for the notation.  Consider the operators acting on a node $n$.  
The diagonal action is simply
\begin{equation}
	E_n^{\scr(i)} \cdot E_n^{\scr(i)}\left|\Phi_{\mathbf q}\right\rangle=C^{\scr{ni}}\left|\Phi_{\mathbf q}\right\rangle
\end{equation}
where $C^{\scr{ni}}$ is the Casimir of the representation associated to the link $ni$. 
The non-diagonal action depends on the pairing at the node $n$. 
We have three cases, depending on the three possible pairings. These are as follows.
Say the node $n$ is in the pairing $(ij),(ef)$, with positive orientation at the two trivalent vertices 
$(i_n,i,j)$ and $(i_n,e,f)$. Then  we have the diagonal double grasping 
\begin{equation}
	E_n^{\scr(i)} \cdot E_n^{\scr(j)}\left|\Phi_{\mathbf{q}}\right\rangle
=\sum_{\mathbf j,\mathbf i}
D^{ij}_{n}\ \Phi({\mathbf j,\mathbf i})\  \left|{\mathbf j},{\mathbf i}\right\rangle.
\label{grasp1}
\end{equation}
while the two possible non-diagonal graspings give 
\begin{equation}
	E_n^{\scr(i)} \cdot E_n^{\scr(k)}\left|\Phi_{\mathbf{q}}\right\rangle
=\sum_{\mathbf j,\mathbf i}\ \Phi({\mathbf j,\mathbf i})
\left(
X^{ik}_{n}\  \left|{\mathbf j},{\mathbf i}\right\rangle
-
Y^{ik}_{n}\  \left|{\mathbf j},(i_n-1),{\mathbf i'}\right\rangle
-
Z^{ik}_{n}\  \left|{\mathbf j},(i_n+1),{\mathbf i'}\right\rangle\right)
\label{grasp2}
\end{equation}
and
\begin{equation}
E_n^{\scr(i)} \cdot E_n^{\scr(l)}\left|\Phi_{\mathbf{q}}\right\rangle 
=\sum_{\mathbf j,\mathbf i}\ \Phi({\mathbf j,\mathbf i})
\left(
X^{il}_{n}\  \left|{\mathbf j},{\mathbf i}\right\rangle
+
Y^{il}_{n}\  \left|{\mathbf j},(i_n-1){\mathbf i'}\right\rangle
+
Z^{il}_{n}\  \left|{\mathbf j},(i_n+1){\mathbf i'}\right\rangle\right).
\label{grasp3}
\end{equation}
and so on cyclically. The quantities $D^{ij}_{n}$, $X^{ij}_{n}$, $Y^{ij}_{n}$ and $Z^{ij}_{n}$ are defined in \cite{I}. Here $\mathbf i'$ indicates the four intertwiners different from $i_n$. 

Inserting the expressions (\ref{psiB}) and (\ref{psiW}) in the expression (\ref{partenza1}) for the propagator, gives the double sum over permutations 
\begin{equation}
{\mathbf G} _{{\mathbf q}\, n,m}^{\scriptscriptstyle ij,kl} 
=\sum_{\sigma'\sigma}\Big[ \sum_{{\mathbf j}, {\mathbf i}} \overline{W(\sigma'({\mathbf j}), \sigma'({\mathbf i}))}  \big(E^{\scriptscriptstyle(i)}_n \cdot E^{\scriptscriptstyle(j)}_n-n_n^{\scriptscriptstyle(i)}\cdot n_n^{\scriptscriptstyle(j)}\big)
\big(E^{\scriptscriptstyle(k)}_m  \cdot E^{\scriptscriptstyle(l)}_m-n_m^{\scriptscriptstyle(k)}\cdot n^{\scriptscriptstyle(l)}_m\big)  \Phi(\sigma({\mathbf j}), \sigma({\mathbf i}))\Big].
\label{partenza3} 
\end{equation} 
The $E$ operators do not change the spin, and the argument at the end of the last section can be repeated.  This time, however, the residual sum over permutations remains, because the operators
are not invariant under it
\begin{equation} 
{\mathbf G} _{{\mathbf q}\, n,m}^{\scriptscriptstyle ij,kl} 
=\sum_{\sigma}\left( \sum_{{\mathbf j}, {\mathbf i}} \overline{W(\sigma({\mathbf j}), \sigma({\mathbf i}))}  \big(E^{\scriptscriptstyle(i)}_n \cdot E^{\scriptscriptstyle(j)}_n-n_n^{\scriptscriptstyle(i)}\cdot n_n^{\scriptscriptstyle(j)}\Big)
\big(E^{\scriptscriptstyle(k)}_m  \cdot E^{\scriptscriptstyle(l)}_m-n_m^{\scriptscriptstyle(k)}\cdot n^{\scriptscriptstyle(l)}_m\big)  \Phi(\sigma({\mathbf j}), \sigma({\mathbf i}))\right).
\label{partenza33} 
\end{equation}
By changing variables, we can move the symmetrization to the operators, hence writing 
\begin{equation} 
{\mathbf G} _{{\mathbf q}\, n,m}^{\scriptscriptstyle ij,kl} 
=\sum_{\sigma} \tilde{\mathbf G} _{{\mathbf q}\, \sigma(n),\sigma(m)}^{\scriptscriptstyle \sigma(i)\sigma(j),\sigma(k)\sigma(l)} 
\end{equation}
where 
\begin{equation} 
\tilde {\mathbf G} _{{\mathbf q}\, n,m}^{\scriptscriptstyle ij,kl} 
=\sum_{{\mathbf j}, {\mathbf i}} \overline{W({\mathbf j},{\mathbf i})}  \big(E^{\scriptscriptstyle(i)}_n \cdot E^{\scriptscriptstyle(j)}_n-n_n^{\scriptscriptstyle(i)}\cdot n_n^{\scriptscriptstyle(j)}\big)
\big(E^{\scriptscriptstyle(k)}_m  \cdot E^{\scriptscriptstyle(l)}_m-n_m^{\scriptscriptstyle(k)}\cdot n^{\scriptscriptstyle(l)}_m\big)  \Phi({\mathbf j}, {\mathbf i}).
\label{partenza4} 
\end{equation}
In other words, we can first compute the propagator with unsymmetrized states and vertex, and then symmetrize the propagator. 

We can now begin the actual calculation of the various terms of the propagator.  It is usuefull
to distinguish three cases: the diagonal--diagonal components  
$\tilde {\mathbf G} _{{\mathbf q}\, n,m}^{\scriptscriptstyle ii,kk}$; the 
diagonal--non-diagonal components
$\tilde {\mathbf G} _{{\mathbf q}\, n,m}^{\scriptscriptstyle ii,kl}$; and the 
non-diagonal--non-diagonal components
$\tilde {\mathbf G} _{{\mathbf q}\, n,m}^{\scriptscriptstyle ij,kl}$, where again different indices are distinct.  Let us considered the three cases separately.

In the diagonal--diagonal case, from the expression of the last section, we have 
 \begin{equation}
\tilde{\mathbf G} _{{\mathbf q}\, n,m}^{\scriptscriptstyle ii,kk}
=\sum_{{\mathbf j}{\mathbf i}}\  
\overline{{W}({\mathbf j},{\mathbf i})}\ 
(C^{ni}-|n_n^{\scriptscriptstyle(i)}|^2)(C^{nk}-|n_m^{\scriptscriptstyle(k)}|^2)\ 
\Phi({\mathbf j},{\mathbf i}) 
\label{partenzadiag}
\end{equation}
As we have seen in \cite{I} the background geometry determines the background link $j^0$
\begin{equation}
	|n_n^{\scriptscriptstyle(i)}|^2=C^2(j^0)=j^0(j^0+1)
\end{equation}
and
\begin{equation}
	C^{ni}=C^2(j^{\scr{ni}}).
\end{equation}
In the large $j^0$ limit we have at leading order
 \begin{equation}
	C^{ni}-|n_n^{\scriptscriptstyle(i)}|^2\approx2j^0 \delta j^{\scr{ni}}\label{gemello diag}
\end{equation}
the propagator components are then
 \begin{equation}
\tilde{\mathbf G}_{{\mathbf q}\, n,m}^{\scriptscriptstyle ii,kk}=4j^2_0\sum_{{\mathbf j},{\mathbf i}}\  
\overline{{W}({\mathbf j},{\mathbf i})}\ 
 \delta j^{\scr{ni}} \, \delta j^{\scr{mk}}\ 
\Phi({\mathbf j},{\mathbf i})
\label{diagfine}
\end{equation}
The sum over permutations is now trivial. It only gives a $5!$ factor that cancels with the same factor in the normalization.  We can therefore drop the tilde from \Ref{diagfine}.

In the diagonal--non-diagonal case, from \Ref{partenza4} we have 
\begin{equation}
	\begin{split}
\tilde	{\mathbf G}_{{\mathbf q}\, n,m}^{\scriptscriptstyle ij,kk}		=
		\sum_{{\mathbf {j}}, {\mathbf {i}}} \ {W}({\mathbf {j}}, {\mathbf {i}})   \big(E^{\scriptscriptstyle(i)}_{n} \cdot E^{\scriptscriptstyle(j)}_{n}-n^{\scriptscriptstyle(i)}_n\!\!\cdot n^{\scriptscriptstyle(j)}_n\big)\big(E^{\scriptscriptstyle(k)}_{m}\cdot  E^{\scriptscriptstyle(k)}_{m}-|n_m^{\scriptscriptstyle(k)}|\big)
				 \Phi({\mathbf {j}}, {\mathbf {i}})
\end{split}
\label{Gsigmandiag}
\end{equation}
now the second operator is diagonal and gives \eqref{gemello diag} at leading order;
the action of the first operator instead gives only one of the three terms \eqref{grasp1}, \eqref{grasp2}, \eqref{grasp3} depending on how the two links $ni$ and $nj$ are paired at the node 
$n$.  The possible results (at leading order) are 
\begin{equation}
	\begin{split}
	\tilde	{\mathbf G}_{{\mathbf q}\, n,m}^{\scriptscriptstyle ij,kk}	=
	\sum_{{\mathbf {j}}, {\mathbf {i}}} \ {W}({\mathbf {j}}, {\mathbf {i}})   
		2j_0\delta j^{\scriptscriptstyle(mk)}  
\left(D^{\scr{(i}\scr{j})}_{n}+\frac{j^2_0}{3}	\right)
				 \Phi({\mathbf {j}}, {\mathbf {i}})
		\end{split}
\label{Gsigmandiag1}
\end{equation}
if the two links are paired. The second term in the parenthesis comes from the fact that the background normals are fixed by the background geometry.  In the large $j^0$  limit 
\begin{equation}
	n_n^{\scriptscriptstyle (i)}\cdot n_n^{\scriptscriptstyle(nj)}\approx-\frac{1}{3}(j_0)^2.
	\label{normali}
\end{equation}
And 
\begin{equation}
	\tilde	{\mathbf G}_{{\mathbf q}\, n,m}^{\scriptscriptstyle ij,kk}	=	\sum_{{\mathbf {j}}, {\mathbf {i}}} \left( \overline{{W}({\mathbf {j}}, {\mathbf {i}})}   
	\left(	
X^{\scr{i}\scr{j}}_{n}+\frac{j^2_0}{3}	\right)
- \overline{{W}({\mathbf j},{\mathbf i',i_n-1)}}\ 
Y^{ij}_{n}   -
\overline{W({\mathbf j},{\mathbf i',i_n+1)}} \  
Z^{ij}_{n}
  \right)  2j_0\delta j^{\scriptscriptstyle mk} \Phi({\mathbf j},{\mathbf i}),
\label{Gsigmandiag2}
\end{equation}
or
\begin{equation}
	\tilde	{\mathbf G}_{{\mathbf q}\, n,m}^{\scriptscriptstyle ij,kk}	=	\sum_{{\mathbf {j}}, {\mathbf {i}}} \left( \overline{{W}({\mathbf {j}}, {\mathbf {i}})}   
	\left(	
X^{\scr{i}\scr{j}}_{n}+\frac{j^2_0}{3}	\right)
+ \overline{{W}({\mathbf j},{\mathbf i',i_n-1)}}\ 
Y^{ij}_{n}   +
\overline{W({\mathbf j},{\mathbf i',i_n+1)}} \  
Z^{ij}_{n}
  \right)  2j_0\delta j^{\scriptscriptstyle mk} \Phi({\mathbf j},{\mathbf i}),
\label{Gsigmandiag3}
\end{equation}
according to orientation, if they are not paired. 

In \eqref{Gsigmandiag2} and \eqref{Gsigmandiag3} the term in $Y$ and $Z$ cancel at the leading order for the following reason.  First, recall from \cite{I} that  $Y$ and $Z$ are equal at leading
order.  The difference between the $Y$-term and the $Z$-term is then only given by the $\pm 1$
in the argument of $W$.  But the dependence of $W$ on $i_n$ is of the form $e^{i\frac{\pi}{2} i_n}$.  Hence (up to subleading terms in the large $j_0$ limit)
\begin{equation}
	W(\mathbf{j},\mathbf i',i_n+1)=-W(\mathbf{j},\mathbf i',i_n-1)\label{w+=w-}
\end{equation}
The different between the two terms is just a sign and they cancel.  Thus we have 
\begin{equation}
	\tilde	{\mathbf G}_{{\mathbf q}\, n,m}^{\scriptscriptstyle ij,kk}	=	\sum_{{\mathbf {j}}, {\mathbf {i}}} \overline{{W}({\mathbf {j}}, {\mathbf {i}})}   
\left(X^{\scr{i}\scr{j}}_{n}+\frac{j^2_0}{3}\right)	
2j_0\delta j^{\scriptscriptstyle mk} \Phi({\mathbf j},{\mathbf i}),
\label{Gsigmandiag2e3fine}
\end{equation}
anytime $ni$ and $nj$ are not paired. 

In the large distance limit we have
\Ref{normali} and
\begin{equation}
D_n^{\scriptscriptstyle ij}-n_n^{\scriptscriptstyle(i)}\cdot n_n^{\scriptscriptstyle(j)}=\frac{C^2(i_n)-C^2(j^{\scriptscriptstyle(ni)})-C^2(j^{\scriptscriptstyle(nj)})}{2}+\frac{1}{3}(j_0)^2.
\label {gemello3}
\end{equation}
Introduce the fluctuations variables  $\delta j_{nj}=j_{nj}-j_{0}$,  and $\delta i_{n}=i_n-i_{0}$ and expand around the background  values $j^0$ and $i^0$. In the large $j_0$ limit (which is also large $i_0$). The dominant term of the \eqref{gemello3} is 
\begin{equation}
	D_n^{\scriptscriptstyle ij}-n^{\scriptscriptstyle(ni)}\cdot n^{\scriptscriptstyle(nj)}=
	\delta i_n \;i_0-\delta j^{\scr{ni}}j_0-\delta j^{\scr{nj}}j_0	.
	\label{ultimo gemello}
\end{equation}
Similarly, using the results of \cite{I},
the $X$ terms are approximated substituting $C^2(j)\approx j^2$ and keeping the dominant terms 
\begin{equation}
X^{\scr{ij}}_n=-\frac{1}{4}\left((i_0)^2+2j_0\;\delta j^{\scr{ni}}+2j_0\;\delta j^{\scr{nj}}-2j_0\;\delta j^{\scr{nf}}-2j_0\;\delta j^{\scr{ne}}+2i_0\;\delta i_n\right)
	\label{xdominant}
\end{equation}
where $nf$ and $ne$ indicate the other two links of the node $n$. 
Recalling that  $i_0=\frac{2}{\sqrt{3}}j_0$, we have that the first term of the sum cancels the
norm of the $n$, leaving 
\begin{equation}
	X^{\scr{ij}}_n+\frac{j_0}3=-\frac14\left(2j_0\;\delta j^{\scr{ni}}+2j_0\;\delta j^{\scr{nj}}-2j_0\;\delta j^{\scr{nf}}-2j_0\;\delta j^{\scr{ne}}+2i_0\;\delta i_n\right)
	\label{xdominant}
\end{equation}
In conclusion, we have for the paired case 
\begin{equation}
\tilde	{\mathbf G}_{{\mathbf q}\, n,m}^{\scriptscriptstyle ij,kk}	=	2j_0^2 \sum_{{\mathbf {j}}, {\mathbf {i}}} \overline{{W}({\mathbf {j}}, {\mathbf {i}})}   
\left(\frac{2}{\sqrt{3}}\delta i_n-\delta j^{\scr{ni}}-\delta j^{\scr{nj}}\right)	\delta j^{\scriptscriptstyle mk}\  \Phi({\mathbf j},{\mathbf i}),
\label{Gsigmandiag2e3fine}
\end{equation}
and for the unpaired one
\begin{equation} 
	\tilde	{\mathbf G}_{{\mathbf q}\, n,m}^{\scriptscriptstyle ij,kk}	= j_0^2 \sum_{{\mathbf {j}}, {\mathbf {i}}} \overline{{W}({\mathbf {j}}, {\mathbf {i}})}   
\left(-\delta j^{\scr{ni}}-\delta j^{\scr{nj}}+\delta j^{\scr{nf}}+\delta j^{\scr{ne}}
-\frac{2}{\sqrt{3}}\delta i_n\right)	
\delta j^{\scriptscriptstyle mk}\  \Phi({\mathbf j},{\mathbf i}).
\label{Gsigmandiag2e3fine}
\end{equation}
Finally, the non-diagonal--non-diagonal case is
\begin{equation}
	\tilde{\mathbf G} _{{\mathbf q}\, n,m}^{\scriptscriptstyle ij,kl}= 
\langle
{W}| 
\big(E^{\scriptscriptstyle(i)}_n  \cdot 
E^{\scriptscriptstyle(j)}_n-n_n^{\scriptscriptstyle(i)}\cdot n^{\scriptscriptstyle(j)}_n\big) 
 \big(E^{\scriptscriptstyle(k)}_m  \cdot 
E^{\scriptscriptstyle(l)}_m-n_m^{\scriptscriptstyle(k)}\cdot n^{\scriptscriptstyle(l)}_m\big)
			\left| \Phi\right\rangle.
\end{equation}
The calculations are clearly the same as above. 

The final result is 
\begin{equation}
	\tilde	{\mathbf G}_{{\mathbf q}\, n,m}^{\scriptscriptstyle ij,kl}	=	j_0^2 \sum_{{\mathbf {j}}, {\mathbf {i}}} \overline{{W}({\mathbf {j}}, {\mathbf {i}})}   
K_n^{ij}K_m^{kl} \  \Phi({\mathbf j},{\mathbf i}),
\label{Gsigmandiagfine}
\end{equation}
where 
\begin{equation}
K_n^{ij}= \frac{2}{\sqrt{3}}\delta i_n-\delta j^{\scr{ni}}-\delta j^{\scr{nj}}
\label{ijpiar}
\end{equation}
if $ni$ and $nj$ are paired at $n$ and 
\begin{equation}
K_n^{ij}= \frac12 \left(-\delta j^{\scr{ni}}-\delta j^{\scr{nj}}+\delta j^{\scr{nf}}+\delta j^{\scr{ne}}
-\frac{2}{\sqrt{3}}\delta i_n\right)	\label{ijunp}
\end{equation}
if they are not; 
while\begin{equation}
K_n^{ii}= 	2 \delta j^{\scriptscriptstyle ni}. 
\label{ii}
\end{equation}

Both the state coefficients $\Phi({\mathbf {j}}, {\mathbf {i}})$ and the vertex coefficients 
$W({\mathbf {j}}, {\mathbf {i}})$ are given by a gaussian in $\delta I_\alpha$. The phases 
in the boundary state cancels with the phase of one of the two terms of $W$, while 
the other term is suppressed for large $j_0$. Thus, \Ref{Gsigmandiagfine} reads 
\begin{equation}
	\tilde	{\mathbf G}_{{\mathbf q}\, n,m}^{\scriptscriptstyle ij,kl}	= j_0^2 \sum_{{\mathbf {j}}, {\mathbf {i}}} 
	e^{-\frac{1}{2j_0}M_{\alpha\beta}\delta I_\alpha \delta I_\beta}
K_n^{ij}K_m^{kl},
\label{Gsigmandiagfine2}
\end{equation}
where $M=A+ij_0G$.  As in \cite{I}, we approximate the sum by a Gaussian integral with
quadratic insertions.  The result of the integral is easily expressed  in terms of the matrix $M^{-1}$ obtained inverting the $15\times15$ covariance matrix $M$, in the 10 spin variables $\delta j_{nm}$ and the five intertwiner variables $\delta i_n$.

The symmetries of the matrix $M^{-1}$ are the same as the symmetries of $M$, and are dictated by the symmetries of the problem. Which ones are these symmetries?  At first sight,
one is tempted to say that $M^{-1}$ must respect the symmetries of the 4-simplex, and therefore 
it must be invariant under any permutation of the five vertices 
$n$. Therefore therefore it can have only seven independent components: 
\begin{eqnarray}
&
M^{-1}_{(ij)(ij)}=c_2,  \ \ \ \ 
M^{-1}_{(ij)(ik)}=c_1, \ \ \ \ 
M^{-1}_{(ij)(kl)}=c_3,
& \no 
&M^{-1}_{ii}=c_4, \ \ \ \ 
M^{-1}_{ij}=c_5, \ \ \ \ 
M^{-1}_{(ij)i}=c_6 \ \ \ \  
M^{-1}_{(ij)k}=c_7.&  \label{sbagliata}
\end{eqnarray}
where different indices are distinct.  The ratio for this being for instance that  $M_{11}$ must be
equal to $M_{22}$ because of the symmetry under the exchange of the vertex 1 and the vertex 2. 
However, this argument is incorrect.

The reason is that the vertex function and the state function are written as a function of 
intertwiner variables $i_n$ which are tied to a given choice of pairing at each node. 
Specifically, we have chosen the pairing $i_1^{(23)(45)}, i_2^{(34)(51)}, i_3^{(45)(12)}, i_4^{(51)(23)}, i_5^{(12)(34)}$.  This choice breaks the symmetry under the permutations of the vertices,
although this is not immediately evident.  To see this, consider for instance the two matrix elements
$M^{-1}_{(12)3}$ and $M^{-1}_{(12)4}$. According to \Ref{sbagliata}, they should be equal (both be equal to $c_7$ by symmetry.  But notice that 1 and 2 are paired at the node 3, while they are not paired at the node 4. 
Therefore the two are not equal under the symmetries of the paired 4-simplex.  To see this more formally, let us indicate explicitly the pairing in which the intertwiner is written by writing $i_n^{(ij)(ef)}$ instead of $i_n$.  Then we see that $M^{-1}_{(12)3}$ is of the form 
$M^{-1}_{(ij)i_n^{(ij)(kl)}}$ while $M^{-1}_{(12)4}$ is of the form $M^{-1}_{(ij)i_n^{(ik)(jl)}}$, which 
makes it obvious that a permutation $ijklm\to i'j'k'l'm'$ cannot transform one into the other, since it cannot undo the fact that the $ij$ indices of the link are paired at the node.  As a consequence, we must for instance replace the last entry of \Ref{sbagliata} by 
\begin{eqnarray}
&M^{-1}_{(ij)i_n^{(ij)(kl)}}=c_7 \ \ \ \  
M^{-1}_{(ij)i_n^{(ik)(jl)}}=c_8.&  \label{giusta}
\end{eqnarray}
and so on.  Thus, the matrix  $M^{-1}$ may in general have a more complicated structure than \Ref{sbagliata}. 

Now, the details of this structure depend on the pairing chosen. 
In fact, there are five possible inequivalent ways of choosing the pairings at the nodes, which do not transform into one another under permutations.  These are illustrated in Figure 1. 
\begin{center}
\begin{figure}[h]
\vskip2cm
\begin{center}
\ifx\JPicScale\undefined\def\JPicScale{0.6}\fi
\psset{unit=\JPicScale mm}
\psset{linewidth=0.3,dotsep=1,hatchwidth=0.3,hatchsep=1.5,shadowsize=1,dimen=middle}
\psset{dotsize=0.7 2.5,dotscale=1 1,fillcolor=black}
\psset{arrowsize=1 2,arrowlength=1,arrowinset=0.25,tbarsize=0.7 5,bracketlength=0.15,rbracketlength=0.15}
\begin{pspicture}(0,0)(50,51)
\psline[linewidth=0.6](19,49)(31,49)
\psline[linewidth=0.2](45,10)(50,27)
\psline[linewidth=0.6](5,10)(15,2)
\psline[linewidth=0.2](19,49)(4,39)
\psline[linewidth=0.2](31,49)(46,39)
\psline[linewidth=0.6](4,39)(0,27)
\psline[linewidth=0.6](46,39)(50,27)
\psline[linewidth=0.2](0,27)(5,10)
\psline[linewidth=0.6](45,10)(35,2)
\psline[linewidth=0.2](15,2)(35,2)
\psline[linewidth=0.2,border=0.3](31,49)(45,10)
\psline[linewidth=0.2,border=0.3](50,27)(15,2)
\psline[linewidth=0.2,border=0.3](35,2)(0,27)
\psline[linewidth=0.2,border=0.3](4,39)(46,39)
\psline[linewidth=0.2,border=0.3](5,10)(19,49)
\rput(30,54){$i^{\scr{(23)(45)}}_{1}$}
\rput(58,35){$i_2^{\scr(34)(51)}$}
\rput(50,3){$i_3^{\scr{(45)(12)}}$}
\rput(1,3){$i_4^{\scr{(51)(23)}}$}
\rput(-6,35){$i_5^{\scr{(12)(34)}}$}
\end{pspicture}\hspace{1.7cm}
\ifx\JPicScale\undefined\def\JPicScale{0.6}\fi
\psset{unit=\JPicScale mm}
\psset{linewidth=0.3,dotsep=1,hatchwidth=0.3,hatchsep=1.5,shadowsize=1,dimen=middle}
\psset{dotsize=0.7 2.5,dotscale=1 1,fillcolor=black}
\psset{arrowsize=1 2,arrowlength=1,arrowinset=0.25,tbarsize=0.7 5,bracketlength=0.15,rbracketlength=0.15}
\begin{pspicture}(0,0)(54,59)
\psline[linewidth=0.6](25,42)(25,55)
\psline[linewidth=0.2](44,9)(49,41)
\psline[linewidth=0.6](17,19)(6,9)
\psline[linewidth=0.2](25,55)(1,41)
\psline[linewidth=0.2](25,55)(49,41)
\psline[linewidth=0.6](1,41)(13,33)
\psline[linewidth=0.6](49,41)(37,33)
\psline[linewidth=0.2](1,41)(6,9)
\psline[linewidth=0.6](44,9)(33,19)
\psline[linewidth=0.2](6,9)(44,9)
\psline[linewidth=0.2,border=0.3](25,42)(33,19)
\psline[linewidth=0.2,border=0.3](37,33)(17,19)
\psline[linewidth=0.2,border=0.3](33,19)(13,33)
\psline[linewidth=0.2,border=0.3](13,33)(37,33)
\psline[linewidth=0.2,border=0.3](17,19)(25,42)
\rput(30,59){$i_1^{\scr(25)(34)}$}
\rput(54,47){$i_2^{\scr(31)(45)}$}
\rput(46,4){$i_3^{\scr{(42)(51)}}$}
\rput(10,4){$i_4^{\scr(53)(12)}$}
\rput(4,47){$i_5^{\scr{(14)(23)}}$}
\end{pspicture}\hspace{1.7cm}
\ifx\JPicScale\undefined\def\JPicScale{0.6}\fi
\psset{unit=\JPicScale mm}
\psset{linewidth=0.3,dotsep=1,hatchwidth=0.3,hatchsep=1.5,shadowsize=1,dimen=middle}
\psset{dotsize=0.7 2.5,dotscale=1 1,fillcolor=black}
\psset{arrowsize=1 2,arrowlength=1,arrowinset=0.25,tbarsize=0.7 5,bracketlength=0.15,rbracketlength=0.15}
\begin{pspicture}(0,0)(50,59)
\psline[linewidth=0.6](25,44)(25,56)
\psline[linewidth=0.2](45,10)(50,27)
\psline[linewidth=0.6](5,10)(15,2)
\psline[linewidth=0.2](25,56)(4,39)
\psline[linewidth=0.2](25,56)(46,39)
\psline[linewidth=0.6](4,39)(0,27)
\psline[linewidth=0.6](46,39)(50,27)
\psline[linewidth=0.2](0,27)(5,10)
\psline[linewidth=0.6](45,10)(35,2)
\psline[linewidth=0.2](15,2)(35,2)
\psline[linewidth=0.2,border=0.3](25,44)(45,10)
\psline[linewidth=0.2,border=0.3](50,27)(15,2)
\psline[linewidth=0.2,border=0.3](35,2)(0,27)
\psline[linewidth=0.2,border=0.3](4,39)(46,39)
\psline[linewidth=0.2,border=0.3](5,10)(25,44)
\rput(58,35){$i_2^{\scr(34)(51)}$}
\rput(49,4){$i_3^{\scr{(45)(12)}}$}
\rput(0,4){$i_4^{\scr{(51)(23)}}$}
\rput(-6,35){$i_5^{\scr{(12)(34)}}$}
\rput(25,59){$i_1^{\scr{(25)(34)}}$}
\end{pspicture}
\end{center}\vspace{5mm}
\begin{center}
\ifx\JPicScale\undefined\def\JPicScale{0.6}\fi
\psset{unit=\JPicScale mm}
\psset{linewidth=0.3,dotsep=1,hatchwidth=0.3,hatchsep=1.5,shadowsize=1,dimen=middle}
\psset{dotsize=0.7 2.5,dotscale=1 1,fillcolor=black}
\psset{arrowsize=1 2,arrowlength=1,arrowinset=0.25,tbarsize=0.7 5,bracketlength=0.15,rbracketlength=0.15}
\begin{pspicture}(0,0)(50,51)
\psline[linewidth=0.6](25,35)(25,47)
\psline[linewidth=0.2](45,10)(50,34)
\psline[linewidth=0.6](5,10)(15,2)
\psline[linewidth=0.2](25,47)(0,34)
\psline[linewidth=0.2](25,47)(50,34)
\psline[linewidth=0.6](0,34)(11,27)
\psline[linewidth=0.6](50,34)(39,27)
\psline[linewidth=0.2](0,34)(5,10)
\psline[linewidth=0.6](45,10)(35,2)
\psline[linewidth=0.2](15,2)(35,2)
\psline[linewidth=0.2,border=0.3](25,35)(45,10)
\psline[linewidth=0.2,border=0.3](39,27)(15,2)
\psline[linewidth=0.2,border=0.3](35,2)(11,27)
\psline[linewidth=0.2,border=0.3](11,27)(39,27)
\psline[linewidth=0.2,border=0.3](5,10)(25,35)
\rput(56,39){$i_2^{\scr(31)(45)}$}
\rput(50,3){$i_3^{\scr{(45)(12)}}$}
\rput(1,3){$i_4^{\scr{(51)(23)}}$}
\rput(0,39){$i_5^{\scr{(14)(23)}}$}
\rput(25,51){$i_1^{\scr(25)(34)}$}
\end{pspicture}
\hspace{2cm}
\ifx\JPicScale\undefined\def\JPicScale{0.6}\fi
\psset{unit=\JPicScale mm}
\psset{linewidth=0.3,dotsep=1,hatchwidth=0.3,hatchsep=1.5,shadowsize=1,dimen=middle}
\psset{dotsize=0.7 2.5,dotscale=1 1,fillcolor=black}
\psset{arrowsize=1 2,arrowlength=1,arrowinset=0.25,tbarsize=0.7 5,bracketlength=0.15,rbracketlength=0.15}
\begin{pspicture}(0,0)(54,48)
\psline[linewidth=0.6](19,44)(31,44)
\psline[linewidth=0.2](45,10)(52,34)
\psline[linewidth=0.6](5,10)(15,2)
\psline[linewidth=0.2](19,44)(-2,34)
\psline[linewidth=0.2](31,44)(52,34)
\psline[linewidth=0.6](-2,34)(9,27)
\psline[linewidth=0.6](52,34)(41,27)
\psline[linewidth=0.2](-2,34)(5,10)
\psline[linewidth=0.6](45,10)(35,2)
\psline[linewidth=0.2](15,2)(35,2)
\psline[linewidth=0.2,border=0.3](31,44)(45,10)
\psline[linewidth=0.2,border=0.3](41,27)(15,2)
\psline[linewidth=0.2,border=0.3](35,2)(9,27)
\psline[linewidth=0.2,border=0.3](9,27)(41,27)
\psline[linewidth=0.2,border=0.3](5,10)(19,44)
\rput(50,3){$i_3^{\scr{(45)(12)}}$}
\rput(1,3){$i_4^{\scr{(51)(23)}}$}
\rput(0,39){$i_5^{\scr{(14)(23)}}$}
\rput(56,39){$i_2^{\scr{(31)(45)}}$}
\rput(30,49){$i_1^{\scr{(23)(45)}}$}
\end{pspicture}
\end{center}
\caption{The five classes of pairings: from the upper left: (10), (5,5), (7,3), (6,4) and (4,3,3).}
\end{figure}
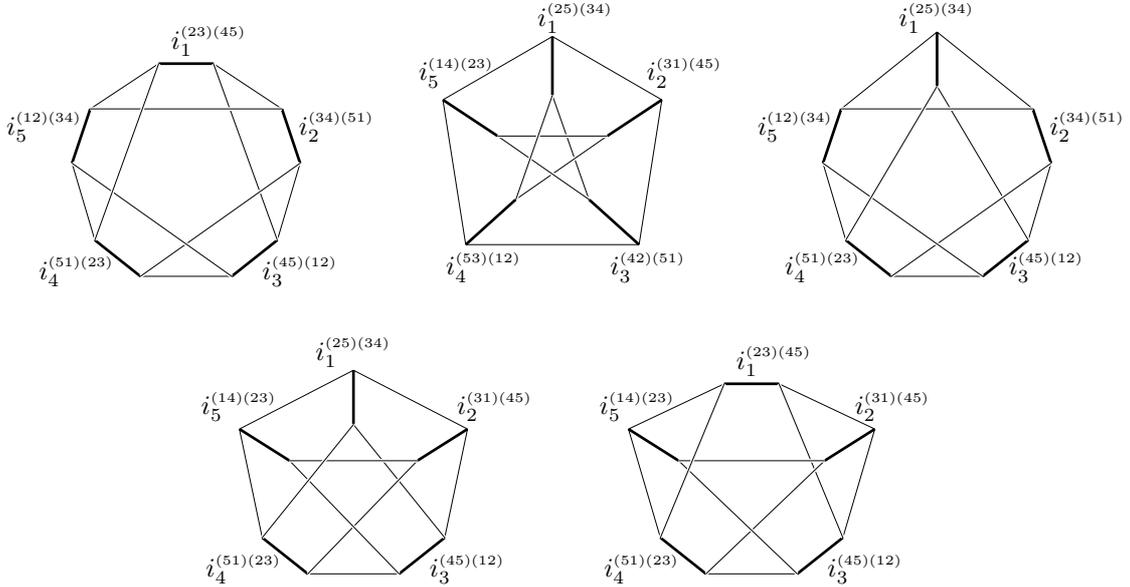
\end{center}
The fact that they cannot be transformed into one another by a permutation can be deduced from the following consideration. In each diagram Figure 1, consider the sequences of links that can be followed without ever crossing an intertwiner.  Observe that in the first case all links are clustered in a single cluster of length 10.  In the second, they are clustered in two diagrams of lenght (5,5), and so on as indicated.   Clearly a permutation cannot change the structure of these clusterings, and therefore these pairing choices cannot be transformed into one another under permutations. The five cases illustrated correspond to the five different 15j Wigner symbols illustrated in \cite{libro15j}.    These five classes define therefore distinct possibilities for the definitionns of vertex and the state.  As here we are not interested in generality, we have just picked one of these: the first case.  Also, since we are not interested in the full generality of an arbitrary gaussian vertex and state, we just assume a particular form, compatible with the symmetries, for the matrix $M^{-1}$. Specifically, we assume that $M^{-1}$ has the form \Ref{sbagliata} with the last entry replaced by 
\Ref{giusta}. That is, we assume the state depends on (at least) eight independent parameters that determine ${\mathbf c}=(c_1,..., c_8)$.    The symmetries of the 4-symplex equivalence class admit a greater number of free parameters, but we do not need the most general possible gaussian state for what follows.  Assuming thus this form for $M^{-1}$, we can then proceed with the calculation of \eqref{Gsigmandiagfine2}.

Each term  of the normalized propagator is a sum of individual elements of the matrix $M^{-1}$.  The overall dependence on $j_0$ is as in the diagonal case, and gives the expected inverse-square dependence.
The normalization factor is 
\begin{equation}
	{\cal N}^{-1}= 
	j_0^2\int d(\delta I_\alpha)\ e^{-\frac{1}{2j_0}M_{\alpha\beta}\delta I_\alpha \delta I_\beta}
\end{equation}
The diagonal-diagonal term gives 
\begin{equation}
	\tilde{\mathbf G} _{{\mathbf q}\, n,m}^{\scriptscriptstyle ii,kk}= {\cal N}
	j_0^2\int d(\delta I_\alpha)\ e^{-\frac{1}{2j_0}M_{\alpha\beta}\delta I_\alpha \delta I_\beta}
	2\delta j_{ni}\ 2\delta j_{mk}\ 
		=
	\frac{4}{j^0}M^{-1}_{\scr({ni})\,(\scr{mk})} =\begin{cases}
	\frac{4}{j^0}c_1 & \textrm{if i=k or i=m},\\
	\frac{4}{j^0}c_3 & \textrm{otherwise}.\end{cases}
	\label{diagfineA}
\end{equation}
In this case  $\tilde {G}$ gives immediately $G$ since the permutation does not
mix $c_1$ and $c_3$ terms. 
	 
Proceeding in the same way for the other cases, we get for the diagonal--non-diagonal term the two cases 
\begin{equation}
\tilde{\mathbf G} _{{\mathbf q}\, n,m}^{\scriptscriptstyle ij,kk}
= 
\frac{1}{j^0}
\big(
-2 M^{-1}_{\scr({mk})\,\scr({ni})}
-2 M^{-1}_{\scr({mk})\,\scr({nj})}
+\frac{4}{\sqrt 3}M^{-1}_{\scr({mk})\,n}\big)
=\begin{cases}
-\frac{4}{j^0}
\big(
c_1 
-\frac{1}{\sqrt 3}c_7\big)& \textrm{if i=k and j=m},
\\
-\frac{4}{j^0}
\big(
c_3 
-\frac{1}{\sqrt 3}c_7\big)& \textrm{if i$\neq$ k and j$\neq$ k,m}, \\
-\frac{2}{j^0}
\big(
c_1+c_3 
-\frac{2}{\sqrt 3}c_7\big)& \textrm{otherwise}.
\end{cases}
\label{ndiagfineA1}
\end{equation}
and 
\begin{eqnarray}
\tilde{\mathbf G} _{{\mathbf q}\, n,m}^{\scriptscriptstyle ij,kk}
&=& 
\frac{1}{j^0}
\big(
 -M^{-1}_{\scr({mk})\,\scr({ni})}
- M^{-1}_{\scr({mk})\,\scr({nj})}
+M^{-1}_{\scr({mk})\,\scr({np})}
+M^{-1}_{\scr({mk})\,\scr({nq})}
-\frac{2}{\sqrt 3}M^{-1}_{\scr({mk})\,\scr({n})}\big)
\\ 
&=& \begin{cases}
\frac{2}{j^0}(-c_1+c_3-\frac{1}{\sqrt{3}} c_8)&\textrm{if i=k and j=m}\\
\frac{2}{j^0}(-c_3+c_1-\frac{1}{\sqrt{3}}c_8)&\textrm{if i$\neq$ k and j$\neq$ k,m}\\
-\frac{2}{\sqrt{3}j^0} c_8 & \textrm{otherwise}
\end{cases} 
\label{ndiagfineA2}
\end{eqnarray}
depending on the pairing of the node $n$.  
For the non-diagonal--non-diagonal terms, we have the three possibilities:  diagonal double grasping on the two nodes
\begin{equation}\begin{split} 
\tilde{\mathbf G} _{{\mathbf q}\, n,m}^{\scriptscriptstyle ij,kl}=
\frac{1}{j_0}\
\big(&\frac43 M^{-1}_{{\scr{m}}\,{\scr{n}}}-\frac{2}{\sqrt 3}\left(M^{-1}_{{\scr{n}}\,{\scr{mk}}}
+M^{-1}_{{\scr{n}}\,{\scr{ml}}}
+M^{-1}_{{\scr{m}}\,{\scr{ni}}}
+M^{-1}_{{\scr{m}}\,{\scr{nj}}}\right)\\
&+
M^{-1}_{{\scr{ni}}\,{\scr{mk}}}+
M^{-1}_{{\scr{ni}}\,{\scr{ml}}}+
M^{-1}_{{\scr{nj}}\,{\scr{mk}}}+
M^{-1}_{{\scr{nj}}\,{\scr{ml}}}\big);
\label{nndiagfineA}
\end{split}
\end{equation}
diagonal double grasping on one node and non-diagonal on the other one
\begin{equation}\begin{split}
&\tilde{{\mathbf G}} _{{\mathbf q}\, n,m}^{\scriptscriptstyle ij,kl}=
\frac{1}{2j_0}
\big(-\frac{4}{3}M^{-1}_{n\;m}	+\\
&-\frac{2}{\sqrt{3}}M^{-1}_{n\,\scr{mk}}-\frac{2}{\sqrt{3}}M^{-1}_{n\,\scr{ml}}
+\frac{2}{\sqrt{3}}M^{-1}_{n\,\scr{mp}}+\frac{2}{\sqrt{3}}M^{-1}_{n\,\scr{mq}}+\\
&+\frac{2}{\sqrt{3}}M^{-1}_{{\scr{ni}}\,m}+ M^{-1}_{{\scr{ni}}\,{\scr{mk}}}+ M^{-1}_{{\scr{ni}}\,{\scr{ml}}}
-M^{-1}_{{\scr{ni}}\,{\scr{mp}}}-M^{-1}_{{\scr{ni}}\,{\scr{mq}}}+\\
&+\frac{2}{\sqrt{3}}M^{-1}_{{\scr{nj}}\,m}+ M^{-1}_{{\scr{nj}}\,{\scr{mk}}}+ M^{-1}_{{\scr{nj}}\,{\scr{ml}}}
-M^{-1}_{{\scr{nj}}\,{\scr{mp}}}-M^{-1}_{{\scr{nj}}\,{\scr{mq}}};
\big),
\end{split}
\end{equation}
and non-diagonal on both nodes 
\begin{equation}\begin{split}
&\tilde{{\mathbf G}} _{{\mathbf q}\, n,m}^{\scriptscriptstyle ij,kl}=
\frac{1}{4j_0}
\big(\frac{4}{3}M^{-1}_{n\;m}	+\\
&+\frac{2}{\sqrt{3}}M^{-1}_{n\,\scr{mk}}+\frac{2}{\sqrt{3}}M^{-1}_{n\,\scr{ml}}
-\frac{2}{\sqrt{3}}M^{-1}_{n\,\scr{mp}}-\frac{2}{\sqrt{3}}M^{-1}_{n\,\scr{mq}}+\\
&+\frac{2}{\sqrt{3}}M^{-1}_{{\scr{ni}}\,m}+ M^{-1}_{{\scr{ni}}\,{\scr{mk}}}+ M^{-1}_{{\scr{ni}}\,{\scr{ml}}}
-M^{-1}_{{\scr{ni}}\,{\scr{mp}}}-M^{-1}_{{\scr{ni}}\,{\scr{mq}}}+\\
&+\frac{2}{\sqrt{3}}M^{-1}_{{\scr{nj}}\,m}+ M^{-1}_{{\scr{nj}}\,{\scr{mk}}}+ M^{-1}_{{\scr{nj}}\,{\scr{ml}}}
-M^{-1}_{{\scr{nj}}\,{\scr{mp}}}-M^{-1}_{{\scr{nj}}\,{\scr{mq}}}+\\
&-\frac{2}{\sqrt{3}} M^{-1}_{{\scr{ne}},m}-M^{-1}_{{\scr{ne}}\,{\scr{mk}}}-M^{-1}_{{\scr{ne}}\,{\scr{ml}}}
+ M^{-1}_{{\scr{ne}}\,{\scr{mp}}}+ M^{-1}_{{\scr{ne}}\,{\scr{mq}}}+\\
&-\frac{2}{\sqrt{3}} M^{-1}_{{\scr{nf}},m}-M^{-1}_{{\scr{nf}}\,{\scr{mk}}}-M^{-1}_{{\scr{nf}}\,{\scr{ml}}}
+ M^{-1}_{{\scr{nf}}\,{\scr{mp}}}+ M^{-1}_{{\scr{nf}}\,{\scr{mq}}}\big) 
\label{nndiagfine1}
\end{split}
\end{equation}
whose expression in terms of the $c$ coefficients in turn depends on pairings.
And so on. 
Notice that the only the six parameters $c_1$, $c_2$, $c_3$ and $c_5$, $c_7$, $c_8$ enter the components of the propagator.  The other two, namely $c_4$ and $c_6$ do not, because we are only looking at the propagator between points on different tetrahedra.   

The last step is to symmetrize the propagator under permutations.   The only terms that change under permutations, at this point, are those due to the pairing. Hence, the only result of a sum over permutation is to combine the two coefficients $c_7$ and $c_8$, which are the only pairing dependent ones.  For instance, a straightforward calculation gives the diagonal--non-diagonal term (which has the peculiarity of not depending on the pairing class)
\begin{eqnarray}
{\mathbf G} _{{\mathbf q}\, n,m}^{\scriptscriptstyle ij,kk}
&=& \sum_\sigma \tilde{\mathbf G} _{{\mathbf q}\, \sigma n,\sigma m}^{\scriptscriptstyle \sigma i\sigma j,\sigma k\sigma k}\label{ndiagfineA2meglio}
\\
&=& 
\begin{cases}
\frac{1}{3j_0}
\big[
{4}(-c_1+c_3)
-{8}
c_1 
   + \frac{2}{\sqrt 3}(c_7-c_8)\big]  & \hspace{3em}   \textrm{if i=k and j=m},\\
\frac{1}{3j_0}
\big[
{4}(-c_3+c_1)
-{4}c_3
+\frac{2}{\sqrt 3}(c_7-{c_8})\big]
& \hspace{3em} \textrm{if i$\neq$ k and j$\neq$ k,m},\\
\frac{1}{3j_0}
\big[
-{2}
(
c_1+c_3
 )
-\frac{1}{\sqrt 3}(c_7-c_8)
\big] &\hspace{3em}  \textrm{otherwise}.\nonumber
\end{cases} 
\end{eqnarray}
It is easy to see that the sum over permutation replaces all terms $c_7$ and $c_8$ with a
term proportional to the linear combination $(c_7-c_8)$.   In conclusion, the propagator depends on the five parameters $c_1,c_2,c_3, c_5, (c_7-c_8)$.   Varying the parameters in the state we can span a five-dimensional space of  tensors $\tilde{\mathbf G} _{{\mathbf q}\, n,m}^{\scriptscriptstyle ij,kl}$.  In conclusion, ${\mathbf G}_{{\mathbf q}\, n,m}^{\scriptscriptstyle ii,kk}$ turns out to be a matrix with the symmetries of the 4-simplex, freely dependending on five arbitrary parameters.  Can this give the same propagator as the linearized theory? 

\section{Comparison with the linearized theory}

The number of components of ${\mathbf G}_{{\mathbf q}\, n,m}^{\scriptscriptstyle ij,kl}$ is large,
and it may seem hard to believe that the five-parameters freedom in the state could be sufficient  to recover the tensorial structure of the linearized propagator.   However, there are two properties of the propagator that strongly constrain it.  First, the symmetrization of the 4-simplex symmetries largely reduce the number of indepedent components.  Second, as proven in  \cite{I}, the propagator satisfies the 
closure relation 
\begin{equation}
\sum_i{\mathbf G}_{{\mathbf q}\, n,m}^{\scriptscriptstyle ij,kl}= 0. 
\label{clos}
\end{equation} 
Let us count the number of free parameters of an arbitrary tensor 
${\mathbf G}_{{\mathbf q}\, n,m}^{\scriptscriptstyle ij,kl}$ satisfying 
these requirements.  Using \Ref{clos}, we can always express a 
term in which any of the four indices $i,j,k,l$ is equal to either $n$
or $m$ as sum of terms not of this kind.  This reduces the independent
terms to, say ${\mathbf G}_{{\mathbf q}\, 1,2}^{\scriptscriptstyle ij,kl}$
where $i,j,k,l=3,4,5$.  A few pictures and a moment of reflection will convince the
reader that the only independent ones of these are 
\begin{equation}\ 
{\mathbf G}_{{\mathbf q}\, n,m}^{\scriptscriptstyle ii,ii}, \ 
{\mathbf G}_{{\mathbf q}\, n,m}^{\scriptscriptstyle ii,kk}, \ 
{\mathbf G}_{{\mathbf q}\, n,m}^{\scriptscriptstyle ij,kk}, \ 
{\mathbf G}_{{\mathbf q}\, n,m}^{\scriptscriptstyle ij,ij}, \ 
{\mathbf G}_{{\mathbf q}\, n,m}^{\scriptscriptstyle ij,ik}.
\label{list}
\end{equation} 
All the other terms can be obtained from these by a permutation of the indices. 
Therefore a tensor with these symmetries depends only on \emph{five} parameters. 
This implies that adjusting the five parameters in the state, we can match
any such tensor, and in particular the propagator. 

This can be checked by an explicit calculation of the propagator of the linearized
theory in the harmonic gauge (on the compatibility of the radial and harmonic gauge, 
see \cite{ClaudioElena}).
The quantity ${\mathbf G}_{{\mathbf q}\, n,m}^{\scriptscriptstyle ij,kl}$ is the propagator
projected in the directions normal to the faces of the tetrahedra.
The 4d linearized graviton propagator is 
\begin{equation}
G_{\mu\nu\rho\sigma}=\frac{1}{2L^2}\;(\delta_{\mu\rho}\delta_{\beta\gamma}+\delta_{\mu\sigma}\delta_{\beta\gamma}-\delta_{\mu\nu}\delta_{\rho\sigma})
\end{equation}
and its projection on the four linear dependent normals to the faces of each tetrahedron reads
 \begin{equation}
G_{nm}^{\scr{ij,kl}}\equiv  G^{\mu\nu\rho\sigma}\
 (n_n^{\scriptscriptstyle(i)})_{\mu} 
 (n_n^{\scriptscriptstyle(j)})_{\nu} 
 (n_m^{\scriptscriptstyle(k)})_{\rho} 
 (n_m^{\scriptscriptstyle(l)})_{\sigma} 
  \label{originale proiettato}
\end{equation}
We need the explicit expressions of the normals; to this aim, fix the coordinate of a four simplex giving the 5 vertices of a 4-simplex fixing the 4d-vectors $e^{\mu}_I$ where $\mu$ is the 4d space index and $I$($I=1,..,5$) is the label of the vertex.   
The easiest way to deal with this 4d geometry is to introduce the bivectors $B^{\mu\nu}_{IJ}$ 
\begin{equation}
	B^{\mu\nu}_{IJ}= e_K^{\mu}\wedge e_L^{\nu}+e_L^{\mu}\wedge e_M^{\nu}+e_M^{\mu}\wedge e_K^{\nu}
\end{equation}
where the indices $IJKLM$ form an even permutation of $1,2,3,4,5$.
If  $t_1$ is the tetrahedron with vertexes $e_2,e_3,e_4,e_5$ and so on cyclically, 
the bivector $B^{\mu\nu}_{nm}$ will be the bivector 
normal to the triangle $t_{nm}$ shared by the tetrahedra $t_n$ and $t_m$. The normal
$n_n^m$  to this triangle, in the 3 surface determined by the tetrahedron $t_n$
is $(n_n^m)^\nu=B^{\mu\nu}_{nm}(t_n)_\mu$, where $(t_n)_\mu$, is the normal to the
tetrahedron. Using this, it is a tedious but straightforward exercise to compute the 
components of the projected linearized propagator. 
Writing the  bimatrix  ${G}_{linearized\, 1,2}^{\scriptscriptstyle ij,kl}=
(G^{kl})^{ij}$, where $ijkl=3,4,5$ we have
\begin{equation}\small
	(G^{kl})^{ij}\sim\frac{1}{512}\begin{pmatrix}
	\begin{pmatrix}
	-16&6&6 \cr
	6&-28&16 \cr
	6&16&-28 \cr
  \end{pmatrix}&\begin{pmatrix}
	6&4&-7 \\
	4&6&-7 \\
	-7&-7&16 
  \end{pmatrix}
  &\begin{pmatrix}
	6&-7&4 \\
	-7&16&-7 \\
	4&-7&6 
  \end{pmatrix} \\
	\begin{pmatrix}
	6&4&-7 \\
	4&6&-7 \\
	-7&-7&16 
  \end{pmatrix}&\begin{pmatrix}
	-28&6&16 \\
	6&-16&6 \\
	16&6&-28 
  \end{pmatrix}&\begin{pmatrix}
	16&-7&-7 \\
	-7&6&4 \\
	-7&4&6 
  \end{pmatrix} \\
	\begin{pmatrix}
	6&-7&4 \\
	-7&16&-7 \\
	4&-7&6 
  \end{pmatrix}&\begin{pmatrix}
	16&-7&-7 \\
	-7&6&4 \\
	-7&4&6 
  \end{pmatrix}&\begin{pmatrix}
	-28&16&6 \\
	16&-28&6 \\
	6&6&-16 
  \end{pmatrix} 
  \end{pmatrix}
  \label{lineare proiettato}
\end{equation}
which displays the equality of the various terms.  The five different components have here values $(-16,6,-28,-7,4)/512$.  A judicious choice of the parameters  $c_1,c_2,c_3, c_5, (c_7-c_8)$ can match these values. 

\section{Conclusion and perspectives}

We have shown that a vertex with an appropriate asymptotic expansion, combined with a suitable  boundary state, can yield the full tensorial structure of the propagator. 

In doing so, we have also learned several lessons.   The main lesson is that the non-commutativity of the angles requires a semiclassical state to have an oscillatory behavior in the intertwiners.  In order to match this behavior, and approximate the semiclassical dynamics, the vertex must have a similar {\em oscillatory dependence on the intertwiners}. (This should not
affect with possible finitness properties of the model  \cite{finiteness}.)    
The second lesson is that the symmetries of the boundary state must be considered with care, if we do not want to loose relevant dynamical information.  Symmetrizing  over the permutation of the vertices is a simple way of achieving a symmetric state without inserting additional unwanted symmetries.  In doing so, however, one must take into account that a choice of pairing breaks the 4-simplex symmetry. 

The most interesting open question, in our opinion, is whether other vertex amplitudes considered
(such as \cite{mike})  and in particular  the vertex amplitude recently studied in \cite{EPR,LS} satisfies the requirements for yielding the correct full tensorial structure of the graviton propagator.  In particular, whether there is an oscillation in the intertwiners. This issue can be addressed analytically, via 
a saddle point analysis of the asymptotic of the new vertex, or numerically, using the technology
developed in \cite{num}.
Some preliminary numerical indications appear to be optimistic \cite{numeric}.  Also, we think that the role of the five inequivalent structures illustrated in Figure 1 deserve to be better understood.


\end{document}